\documentclass{desyproc}

\usepackage{array}
\usepackage{amsmath}
\usepackage{feynmp}
\usepackage{framed,feynmp,graphicx}
\usepackage{mathrsfs}
\usepackage{slashed}
\usepackage{subcaption}
\usepackage{units}

\newcommand{\beq}{\begin{equation}}
\newcommand{\eeq}{\end{equation}}
\newcommand{\bea}{\begin{eqnarray}}
\newcommand{\eea}{\end{eqnarray}}

\newcommand{\A}{\mathcal{A}}
\newcommand{\F}{\mathcal{F}}
\renewcommand{\O}{\mathbf{O}}
\newcommand{\T}{\mathbf{T}}
\newcommand{\U}{\mathbf{U}}
\newcommand{\V}{\mathbf{V}}
\newcommand{\at}{\tilde{a}}
\newcommand{\Bt}{{\tilde{B}}}
\newcommand{\Wt}{{\tilde{W}}}

\newcommand{\LL}{\mathscr{L}}

\newcommand{\WWd}{W_{\mu\nu}}

\newcommand{\de}{\partial}

\newcommand{\DLR}{\mathbf{D}}

\newcommand{\ETmiss}{\slashed{E}_T}

\newcommand{\cl}{\%\,\,  \text{C.L.}}

\DeclareMathOperator{\Tr}{Tr}
\newcommand{\tr}{\Tr}
\DeclareMathOperator{\Br}{Br}

\begin{document}
%------------------------------------
\title{ALPs EFT \& Collider Signatures}

%for single authors the superscripts are optional
\author{{\slshape Roc\'io del Rey Bajo}\\[1ex]
Universidad Aut\'onoma de Madrid \& Instituto de F\'isica Te\'orica, Madrid, Spain}

% if the proceedings are available online (e.g. at Indico)
% please enter the contribution ID or file_name below for the DOI
%\contribID{32}
\contribID{familyname\_firstname}

% TO THE CONFERENCE EDITORS: 
% please update the following information      
% before sending the template to the authors
\confID{13889}  % if the conference is on Indico uncomment this line
\desyproc{DESY-PROC-2017-XX}
\acronym{Patras 2017} % if you want the Acronym in the page footer uncomment this line
\doi  % if there is an online version we will register DOIs

\maketitle

\begin{abstract}
We explore the leading effective interactions between the Standard Model and a generic singlet CP-odd (pseudo)Goldstone boson in two frameworks for electroweak symmetry breaking: linear and non-linear realizations, determining the  basis of leading effective operators for the latter. New bounds are obtained and prospects of signals at colliders are explored. Mono-$Z$, mono-$W$, and $aW\gamma$ are signals expected in both frameworks while non-standard Higgs decays and mono-$h$ signals may point to non-linear EWSB realizations.

\end{abstract}

\vspace{-3mm}
\section{Motivations}
The Higgs discovery  has set spin zero particles
%, proposed in solutions to major problems in particle physics, 
in the spotlight of searches for physics beyond the Standard Model.
 One example is the strong CP problem, whose paradigmatic solution predicts a  (pseudo)Nambu-Goldstone boson (pNGB): the axion. In a QCD-like theory its mass $m_a$ arises because of an explicit breaking of the chiral $U(1)_{PQ}$ symmetry by instantons at a scale $\Lambda$. It is related to the PQ-breaking scale $f_a$ (in a one quark approximation, of mass $m_q$):~\cite{Choi:1986zw} 
 \beq
 m_a^2 f_a^2 = \Lambda^4/\left(1+\Lambda^4/(2m_q <\bar{\psi}\psi>)\right)\,.
 \eeq
In the SM, $\Lambda=\Lambda_{QCD} \gg m_q$, giving the relationship typical of invisible axion models: $m_a f_a \sim m_\pi f_\pi$. With an alternative confining gauge group with a scale $\Lambda'\gg \Lambda_{\text{QCD}}$ this relationship may be modified, e.g. $m_a f_a \sim \Lambda'^2$. In models of this kind (eg. Ref.~\cite{Gherghetta:2016fhp}) axions may have simultaneously a low axion scale $f_a \sim\mathcal{O}( 10\,\text{TeV})$ and a high mass $m_a \sim \mathcal{O}(100 \,\text{GeV})$. 

The nature of the Higgs itself also raises a quandary, as the electroweak (EW) hierarchy problem remains unsolved. The lightness of the Higgs  may result from its being a pNGB of a global symmetry~\cite{Kaplan:1983fs}, spontaneously broken by strong dynamics at a scale $\Lambda_s\gg v$, where $v$ denotes the EW scale, as typically arises in scenarios where electroweak symmetry breaking (EWSB) is non-linearly realised. Much as the interactions of QCD pions are weighted down by the pion decay constant,
% and described by an effective field theory (EFT) with a derivative ordering,  
those of the EW (pseudo)Goldstone Bosons -- the longitudinal components of the $W^\pm$ and $Z$ plus the $h$-- will be weighted 
down by a scale $f$ ($\Lambda_s\leq 4\pi f $)\cite{Manohar:1983md}. 
An effective field theory (EFT) approach allows to avoid the specificities of particular models.

We formulate the leading CP-invariant effective couplings of an extra CP-odd pNGB singlet scalar to SM fields, which must be purely derivative couplings when its mass is neglected. While the dominant ALP interactions in the linear --often called SMEFT-- expansion have been formulated long ago~\cite{Georgi:1986df}, the analogous for the non-linear --often called Higgs EFT (HEFT)-- regime is developed here. 

Up to now, phenomenological analyses concentrated on ALP couplings to  photons, 
gluons and fermions, which dominate at low energies and determine 
astrophysical constraints for light ALPs. However, ALPs may show up first 
at colliders
and the $SU(2)\times U(1)$ invariant 
formulation of their interactions provides new channels involving the EW gauge bosons  and the Higgs. 

\vspace{-4mm}
\section{Linear and Chiral Lagrangians}
\vspace{-3mm}
  Focusing on interactions involving only one ALP, 
  the effective linear ALP Lagrangian including leading (LO) and next-to-leading-order (NLO) interactions was determined long ago~\cite{Georgi:1986df}. The complete basis for ALP interactions with the SM fields at NLO consists of four operators:
\bea
 \A_{\tilde{X}} =-X_{\mu\nu}^a \tilde{X}^{a\mu\nu}\frac{a}{f_a}\quad (X=B,\,W,\,G\,),\qquad \text{and} \qquad
\O_{a\Phi}=  i (\Phi^\dag\overleftrightarrow{D}_\mu\Phi)\frac{\de^\mu a}{f_a}\,, \label{linear_operators}
\eea
where $\A_{\tilde{X}}$ describe ALP couplings to gauge bosons and $\O_{a\Phi}$ induces a two-point function contribution that can be traded for a fermionic vertex $\sim c_{a\Phi}\,( \bar{\psi} \gamma_\mu \gamma_5 \psi)\, \partial^\mu a / f_a$. In this context $\Phi = \frac{v+h}{\sqrt{2}}\U(x)\begin{pmatrix} 0 & 1 \end{pmatrix}^T$, where $\U(x)\equiv e^{i\sigma_a \pi^a(x)/v}$ encodes the $W^\pm$ and $Z$ longitudinal components, denoted by  $\vec{\pi}(x)$.

%In the SMEFT the physical Higgs and the electroweak pGBs, denoted by  $\pi^{a}(x)$, are part of the same object: $\Phi = \frac{v+h}{\sqrt{2}}\U(x)\begin{pmatrix} 0 \\ 1 \end{pmatrix}$, where $\U(x)\equiv e^{i\sigma_a \pi^a(x)/v}$.

In non-linear setups~\cite{Contino:2010rs,Alonso:2012px,Brivio:2016fzo}, the physical Higgs may no longer behave as an exact EW doublet at low energies. It can be treated effectively as a generic scalar singlet with arbitrary couplings by replacing the SM $(v+h)$ dependence
% on $(v+h)$ 
by $\F(h)= 1 + 2a\frac{h}{v}+ b \left(\frac{h}{v}\right)^2 + \dots\,$, whose interactions are not necessarily correlated with those of the EW-``pions" in $\U(x)$.~\footnote{Note that the ``pions" in $\U(x)$ are suppressed by $v$, where the natural GB weight is in fact the scale $f$; this encodes the fine-tuning affecting these models.}
%, $\pi^a(x)$. In non-linear setups 
In non-linear EFTs $\F(h)$ and $\U(x)$ are independent building blocks.~\footnote{Two $SU(2)_L$ covariant objects are used as building blocks containing $\U(x)$: $\V_\mu(x)\equiv \left(\DLR_\mu\U(x)\right)\U(x)^\dag$ and $\,\T(x)\equiv \U(x)\sigma_3\U(x)^\dag$. }
%In addition, the interactions of $h$ are not necessarily correlated with those of the $W^\pm$ and $Z$ longitudinal components, denoted by  $\pi(x)$ in the unitary GB  matrix $\U(x)\equiv e^{i\sigma_a \pi^a(x)/v}$. In non-linear setups $h$ and $\U(x)$ are no longer parts of the same object $\Phi$, but now introduced intependentlty:  in $\F(h)$ and
 We generalize the effective chiral Lagrangian  to include ALP insertions,

\beq
\LL_\text{eff}^\text{chiral}= \LL^{\text{LO}}_{\text{HEFT}} +\frac{1}{2} (\de_\mu a)(\de^\mu a)+c_{2D}\A_{2D}(h)+\, \delta\LL_a^\text{bosonic}\,.
\label{Lchiral}
\eeq
Now, the LO Lagrangian includes the usual HEFT LO terms plus the ALP kinetic term and
\beq
\A_{2D}(h)=iv^2\tr[\T\V_\mu]\de^\mu\frac{a}{f_a}\F_{2D}(h)\,.
\label{Eq:A2D}
\eeq
If EWSB is non-linearly realised $\A_{2D}(h)$ may provide the dominant signals, as it appears singled out at the LO in the chiral expansion.  
Apart from contributing to the $Z^\mu \partial^\mu a$ two-point function, alike to $\O_{a\Phi}$, it additionally gives rise to
$(Z_\mu \de^\mu a) h^n,\, n\geq 1$ couplings, which are {\it not} redefined away as in the non-linear case. The reason is that  the    functional dependence on $h$ of  $\F_i(h)$ differs generically from that characteristic of the linear regime, in powers of $(v+h)^2$. 
 
We find \cite{Brivio:2017ije} that the NLO Lagrangian, $\delta\LL_a^\text{bosonic}=\sum_{\tilde{X}} c_{\tilde{X}}\A_{\tilde{X}}+\sum_{1}^{17}c_i\A_i\,$, consists of 20 independent bosonic structures:
three $\A_{\tilde{X}}$, the same as in Eq.~\eqref{linear_operators} plus 17 operators $\A_i$ (see Ref.~\cite{Brivio:2017ije}) amongst which the ones testable with present and high luminosity LHC data are
\bea
\A_3(h)=\frac{1}{4\pi} B_{\mu\nu}\de^\mu \frac{a}{f_a}\de^\nu \,\F_3(h)\qquad \text{and} \ \qquad \A_6(h)  = \frac{1}{4\pi} \tr[\T[\WWd,\V^\mu]]\de^\nu \frac{a}{f_a} \,\F_6(h) \,.
\eea
\vspace{-4mm}
\section{ALP phenomenology}
\vspace{-3mm}
We showcase some salient features of the effective ALP interactions varying one $c_{\tilde{X}}$ or $c_i$ coefficient  at a time, which allows to single out the impact of each effective operator. The ALPs are considered stable on collider scales~\cite{Brivio:2017ije}, so the bounds LHC presented are valid for $m_a \lesssim \mathcal{O}(\text{MeV})$.

\vspace{-3mm}
\subsection{Signatures present in both linear and non-linear expansions}
\vspace{-1mm}
The ALP coupling to photons has been extensively tested, and is very strongly constrained: Beam Dump experiments~\cite{Bjorken:1984mt} (astrophysical observations~\cite{Raffelt:2006cw}) 
  enforce $|c_\theta^2c_\Bt + s_\theta^2 c_\Wt|$ to cancel to one part 
in $10^{3}$ ($10^{8}$) for $m_a=\unit[1]{MeV\, (keV)}$. To account for these strong constraints we have performed the analyses imposing $g_{a\gamma\gamma}=0\, \longrightarrow \, c_{\tilde{B}} = -t_\theta^2 c_{\tilde{W}},\, $ effectively reducing the number of parameters by one. However, when considering the complete basis of operators from Eqs.~\eqref{linear_operators} or~\eqref{Lchiral}, this coupling
is generated simultaneously to set of couplings to the EW gauge bosons:
\vspace{-.3cm}
\begin{figure}[h]
\centerline{\includegraphics[width=.8\textwidth]{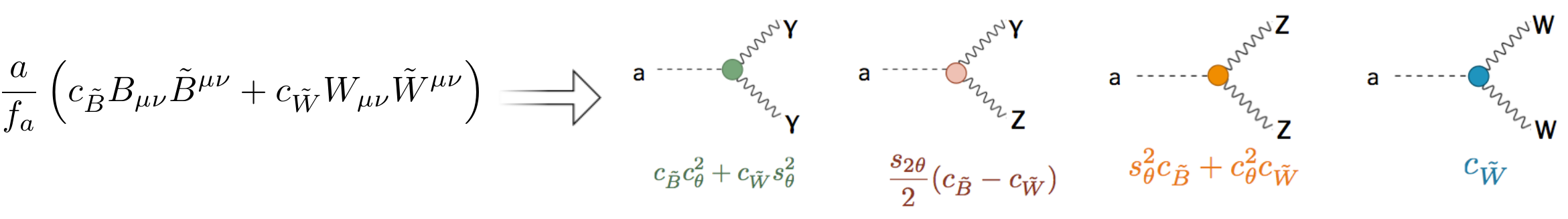}}
\end{figure}
\vspace{-.5cm}
\newline Processes where ALPs are produced through their couplings to gauge bosons allow to test directions of the $\{ c_{\tilde{B}},\,c_{\tilde{W}} \} $ parameter space other than the combination tested through $g_{a\gamma\gamma}$.

\begin{figure}[t]
\centering

\subcaptionbox{\it \small Constraints on $c_{\tilde{B}}/f_a$ and $c_{\tilde{W}}/f_a$ from  the tree-level bounds on
$g_{a\gamma\gamma}$ and $g_{aZ\gamma}$.
 \label{fig:cBcW}
}%
  [.46\linewidth]{\includegraphics[width=.38\textwidth]{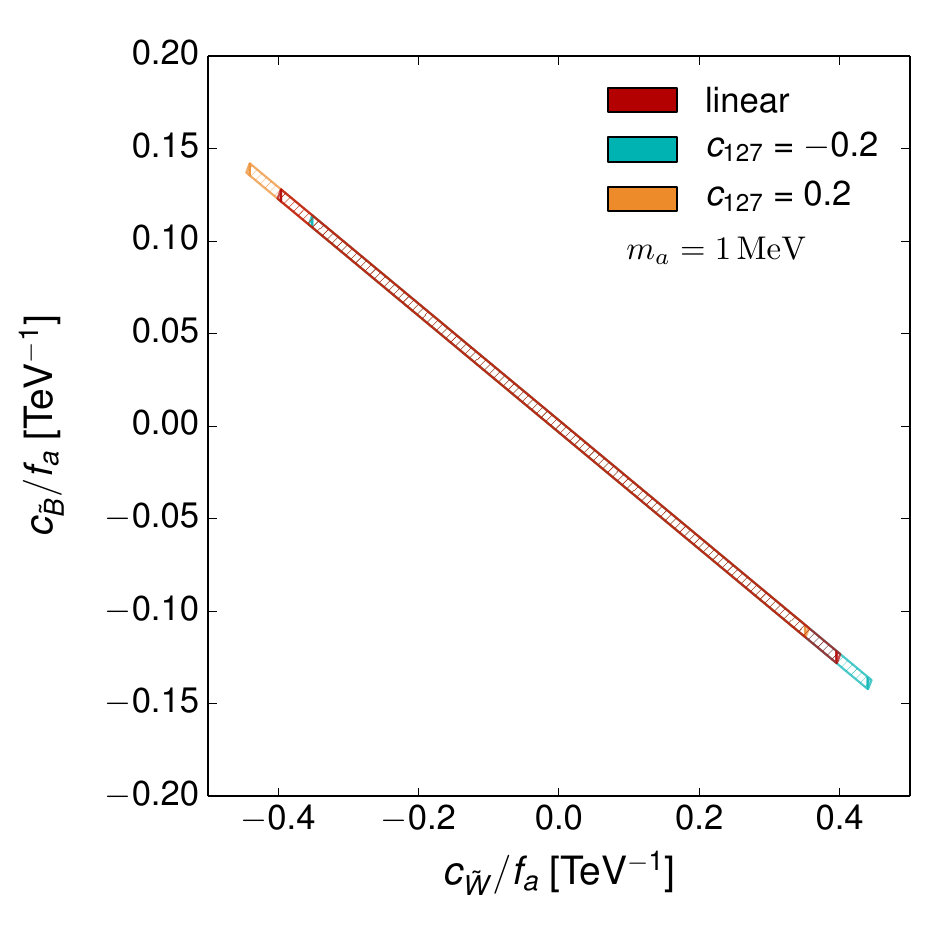}} 
  \hspace{.5cm}
\subcaptionbox{\it \small Constraints from the studies on tree-level ALP couplings in this work.
 \label{fig:summary}}
  [.46\linewidth]{\includegraphics[width=.49\textwidth]{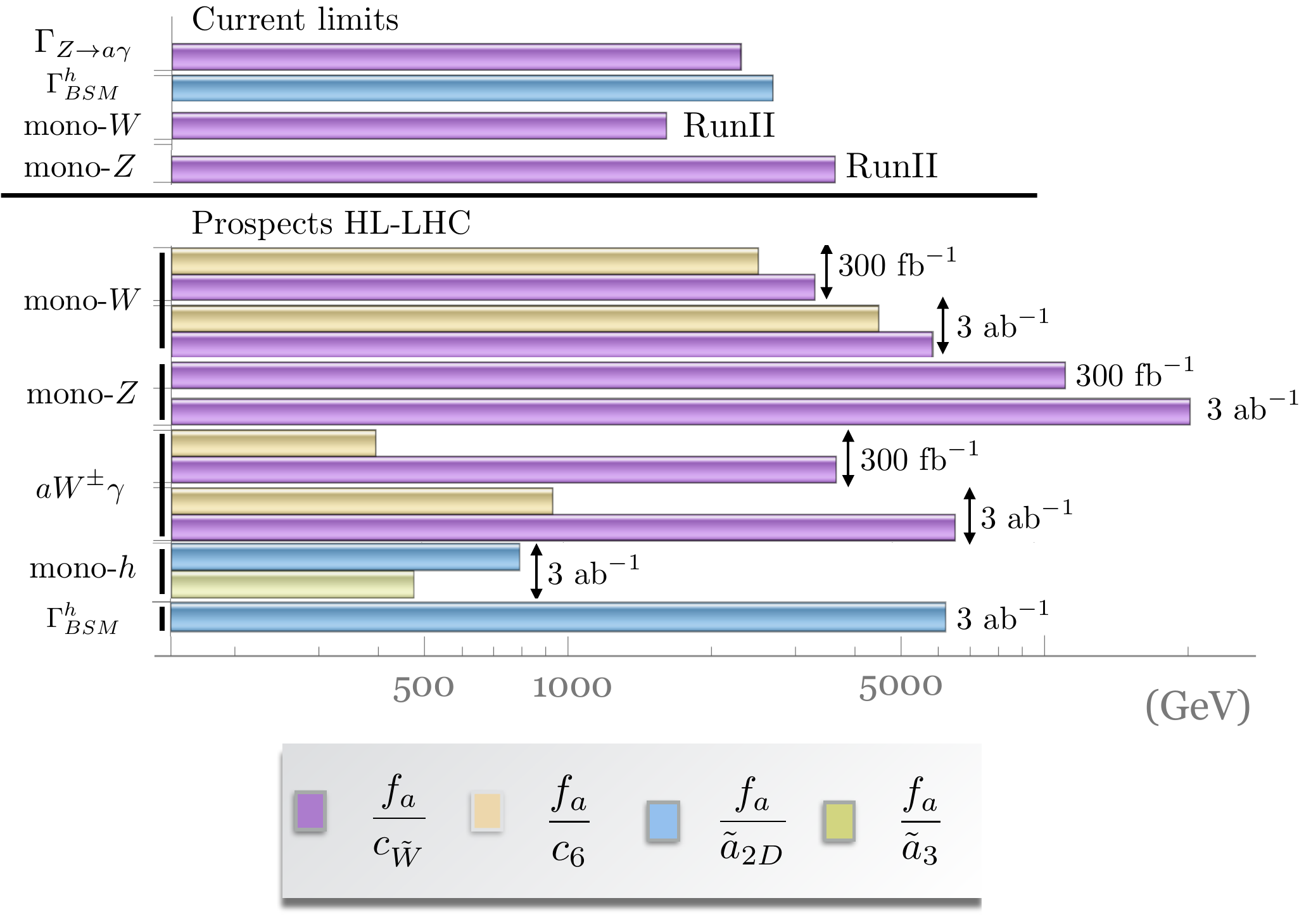}}
\end{figure}

\vspace{0.1cm}
\noindent\textbf{$\,\bullet\,\,$ \underline{ALP coupling to Z-photon:}}\quad 
In the linear expansion, the coupling $   - \frac{1}{4}g_{aZ\gamma} \,a\,Z_{\mu\nu}\tilde{F}^{\mu\nu}$ takes the form
$ g_{aZ\gamma} =4\sin({2\theta})(c_{\tilde{W}}-c_{\tilde{B}}) / f_a$ (with additional operators contributing in the non-linear case), and can be constrained through conservative bounds on $Z\to a\gamma$ from the uncertainty on the $Z$ boson width. Combined with $g_{a\gamma\gamma}$ bounds, $c_{\tilde{B}}$ and $c_{\tilde{W}}$ take values within a limited area, see Fig.~\ref{fig:cBcW}. Imposing $g_{a\gamma\gamma}=0$
leads to $|f_a/c_{\tilde{W}}|> \unit[2.4]{TeV}$.

\vspace{0.1cm}
\noindent\textbf{$\,\bullet\,\,$ \underline{Mono-$Z$ and mono-$W$ signatures:}}\quad The ALP can be produced in association with a $W$ or a $Z$ boson, 
and escape the LHC 
detectors as missing energy.
 Mono-$Z$ signals can be tested reinterpreting the CMS 
$Z + \ETmiss$ search~\cite{CMS:2016yfc} and, as illustrated in Fig.~\ref{fig:summary}, this allows to impose the strongest limit on $c_\Wt$:  $|f_a/c_\Wt |> \unit[3.8]{TeV}$ with Run II data, and 
 up to $|f_a/c_\Wt| \gtrsim 21$ TeV if no signal is found in the high-luminosity (HL-LHC) phase, here computed as $\unit[3]{ab^{-1}}$ of data from $\sqrt{s}=\unit[13]{TeV}$ collisions if nothing is found. 
%~\footnote{We denote by HL-LHC the high luminosity phase of LHC; the analyses here having been performed with $\unit[3]{ab^{-1}}$ of data taken in $\sqrt{s}=\unit[13]{TeV}$ collisions.}
 Future mono-$W$ searches, analysed from the ATLAS search for $W'\to\ell + \ETmiss$ ~\cite{ATLAS-CONF-2015-063} 
may uncover $\mathcal{A}_{6}$, a signal of non-linearity: it will be tested up to $|f_a / c_6| < \unit[4.5]{TeV}$ at HL-LHC. 

\vspace{0.1cm}
\noindent\textbf{$\,\bullet\,\,$ \underline{Associated production $pp\rightarrow W^\pm a \gamma$:}}\quad 
Mono-$W$ and $aW\gamma$ both depend on similar combinations of coefficients: correlation effects are thus {\it a priori} expected amongst them, and their combined analysis should allow to disentangle the measurement of $c_{\tilde{W}}$   from that of $c_6$. Analysed independently, 
  $aW\gamma$
 will reach $|f_a / c_\Wt| \lesssim  \unit[6.8]{\,TeV}$
and $|f_a / c_6| \lesssim\unit[0.8]{\,TeV}$ at HL-LHC.

\vspace{-2mm}
\subsection{Distinctive non-linear signatures: Higgs signatures}
\vspace{-2mm}
ALP-Higgs couplings are an interesting class of new signals which may point to non-linear realizations of EWSB where $aZh^{n}$ vertices are expected at LO, since they do not appear in the linear expansion below NNLO.
The bounds are computed for one operator at a time (for $\A_{2D}$, $\A_3$ and $\A_{10}$ from Ref.~\cite{Brivio:2017ije}), but a combined analysis
may allow to disentangle them. 

\vspace{0.1cm}
\noindent\textbf{$\,\bullet\,\,$ \underline{Non-standard higgs decays $\Gamma^h_{BSM}$:}}\quad 
Current constraints from ATLAS and CMS global fits of   $7$ and $8$ TeV data  to Higgs signal strengths yield~\cite{Khachatryan:2016vau}
$\Br (h\to {\rm BSM})   \leq 0.34 \,\,\, 
(95\cl)\,,$
with 
$\Gamma^h_\text{BSM}\simeq \Gamma_{h \to aZ}$. Interpreting this bound in terms of the LO operator $\A_{2D}$ gives
%.
$
|{f_a}/{\tilde{a}_{2D}}| \gtrsim \unit[2.78]{TeV}\,  \quad \text{for}\quad  m_a \lesssim \unit[34]{GeV}\,,
$ 
where $\at_i \equiv c_i a_i$.
These limits are expected to improve significantly at HL-LHC, which will be able to test up to $|f_a/\at_{2D} |\lesssim \unit[6]{TeV}$.

\vspace{1mm}
\noindent \textbf{$\,\bullet\,\,$ \underline{Mono-$h$:}}\quad 
The  $\ETmiss$ spectrum of $ p p \to a h \,(h \to 4\ell)$ is considered at 13 TeV LHC. Although this search is in principle sensitive to both $\A_{2D}$ and $\A_3$, it is not competitive with Br($h\to {\rm BSM})$ in constraining the former. On the other hand, it is more sensitive to the presence of $\at_3$ (see Fig.~\ref{fig:summary}) and may therefore provide valuable, 
complementary information.

\vspace{-3mm}
\section{ Conclusions}
\vspace{-3mm}

In summary, a general approach to ALP interactions with the SM fields, considering complete effective Lagrangians that include all $SU(3)\times SU(2) \times U(1)$ invariant operators might suggest axion searches complementary to the traditional ones performed until now. Through their coupling to EW gauge bosons we show that ALPs stable at LHC, can be tested via mono-$W$, mono-$Z$, $pp\to aW\gamma$ signatures, with present data (HL-LHC) testing the operator coefficients to $f_a / c_i \sim \mathcal{O}(\unit[1]{TeV})\,\left( \mathcal{O}(\unit[20]{TeV})\, \right)$ for some operators.
%and HL-LHC reaching up to $f_a / c_i \sim\mathcal{O}(\unit[20]{TeV})$ for some operators. 
Furthermore, we have determined the chiral effective Lagrangian for non-linear EWSB. New additional couplings are expected, one of the most promising signals being ALP couplings to the Higgs particle. We show that LHC data for mono-$h$ and non-standard $h$ decays is able to test $aZh$ and $a\gamma h$ interactions, which are not expected in linear expansions below NNLO, and are thus smoking guns for non-linearity.

\begin{footnotesize}

\vspace{-2mm}
\providecommand{\href}[2]{#2}\begingroup\raggedright\endgroup
\end{footnotesize}

% ****************************************************************************
% END OF BIBLIOGRAPHY AREA
% ****************************************************************************

\end{document}